\documentclass[conference]{IEEEtran}
\usepackage[T1]{fontenc}
\usepackage{amsmath,booktabs,tabularx,enumitem,tikz,etoolbox,url}
\usetikzlibrary{arrows.meta,positioning,calc}
\patchcmd{\abstract}{\textit{\abstractname}---\relax}{\textit{\abstractname}\space-\space\relax}{}{}

\newcommand{\na}{-}
\title{Does AI Help Cyber Attackers or Defenders?\\Evidence from Nonpublic Vulnerabilities and Subsequent Attacks}
\author{
\IEEEauthorblockN{Tobias Heldt\IEEEauthorrefmark{1}, Matt Turk\IEEEauthorrefmark{2},
Christoph Landolt\IEEEauthorrefmark{3}, Mario Fritz\IEEEauthorrefmark{3}}
\IEEEauthorblockA{\IEEEauthorrefmark{1}XOR\quad\IEEEauthorrefmark{2}Protege\quad\IEEEauthorrefmark{3}CISPA Helmholtz Center for Information Security}
}

\begin{document}
\maketitle
\begin{abstract}
The release decision for frontier AI systems increasingly relies on cyber capability benchmarks, yet public vulnerability benchmarks can expose agents to previously published advisories, exploits, and fixes, making it difficult to distinguish prior exposure from capability on unseen vulnerabilities. We evaluate open-weight and proprietary AI models on exploit generation, vulnerability repair and subsequent attacks in five nonpublic software environments, including vulnerabilities we privately disclosed while they remained unpatched. Researcher-developed and reviewed deterministic graders, not LLM judges, determine task scores. Comparisons with 209 disclosed vulnerabilities and cryptographic challenges reveal substantial variation across systems and vulnerability types. Repair scores exceed attack scores in two nonpublic environments and fall below them in three. Passing an initial security test is also insufficient: another exploit succeeds in 92 of 524 non-independent defender test intervals after the initial exploit is stopped. These results motivate vulnerability-specific attack-repair comparisons and subsequent resistance tests.
\end{abstract}

\section{Introduction}
AI release decisions depend partly on whether new systems help attackers more than defenders~\cite{oracle}. Cybersecurity benchmarks address this question by asking agents to exploit or repair vulnerable software. However, success can involve actions outside the intended task. In the July 2026 Hugging Face incident, OpenAI agents operating with reduced safeguards coordinated through unauthorized channels, exploited previously unknown vulnerabilities to bypass network restrictions, and compromised OpenAI research infrastructure and Hugging Face servers while seeking benchmark solutions. Agents also copied private evaluation data into a public dataset~\cite{evaluationincident}. This demonstrated offensive capability and a failure of evaluation containment, not merely answer lookup. Even within the intended task, a repair that stops the original exploit may fail against a subsequent attack.

We study five software environments, each comprising a target program, execution setup and evaluation tests. Four concern flaws we investigated; one is a constructed protocol variant. The collection includes vulnerabilities we privately disclosed while they remained unpatched, without public flaw-specific advisories or fixes at evaluation time. We evaluate multiple agent systems, including GPT-6 Sol, and compare their outcomes separately with paired tasks on 209 disclosed memory-safety vulnerabilities and cryptographic challenges.

An \emph{agent system} combines a model with a \emph{harness}, the software that provides tools and executes the model's actions. The attacker submits an exploit; the defender submits a repair. We compare their success rates or task scores and then examine whether further attacks defeat defensive work. Throughout, ``help'' means autonomous task performance, not an improvement in human performance.

Three questions organize the study:
\begin{enumerate}[leftmargin=*,label=\textbf{Q\arabic*.},itemsep=1pt]
\item \textbf{Attack versus repair:} Which systems and vulnerability groups show stronger attack or repair performance?
\item \textbf{Information and verification:} Do more informative instructions and self-review improve repairs, and do agents' checks agree with external tests?
\item \textbf{Repair robustness:} Can further attacks succeed after a defender modifies the software or stops the original exploit?
\end{enumerate}
We show that attack-repair differences change direction across environments, systems and weakness groups. More informative instructions improve selected repairs, while subsequent tests expose failures that initial checks miss. We contribute subgroup comparisons, an experiment on information and self-review, and an analysis of attacks following defensive changes. Figure~\ref{fig:design} maps these experiments; Section~\ref{sec:measurement} defines a common notation for their distinct outcomes.

\section{Related Work}
\paragraph{Measuring security capability}
Happe and Cito review 19 studies and recommend stronger testbeds, baselines and metrics~\cite{benchmarking}. Landolt et al. connect the attacker-defender capability gap to release timing~\cite{oracle}. We provide empirical comparisons by system and vulnerability group to inform this question.
\paragraph{Exposure and shortcut success}
NYU CTF Bench and Cybench provide executable capture-the-flag (CTF) tasks~\cite{nyuctf,cybench}. Milner et al. inspect agent action logs to distinguish exploitation from exposed answers and external solution lookup~\cite{traceprovenance}. CTFusion uses live competitions to reduce exposure to existing solutions~\cite{ctfusion}; AutoBaxBuilder generates code-security tasks and validates their functionality and security~\cite{autobax}. We complement these approaches with vulnerabilities whose flaw-specific solutions were not public, and evaluate attack, repair and subsequent attacks separately.
\paragraph{Reproduced vulnerability benchmarks}
CyberGym evaluates proof-of-concept generation from vulnerability descriptions and repositories~\cite{cybergym}; ARVO supplies reproducible vulnerable and fixed states~\cite{arvo}. We add tasks with researcher-attested absence of public flaw-specific solutions.

\paragraph{Repair and both-role evaluation}
SWE-bench evaluates repository-issue resolution~\cite{swebench}; SEC-bench covers exploit generation and vulnerability patching~\cite{secbench}; BountyBench compares Detect, Exploit and Patch tasks~\cite{bountybench}. We examine subgroup differences and subsequent attacks on nonpublic tasks.

\section{Study Overview}
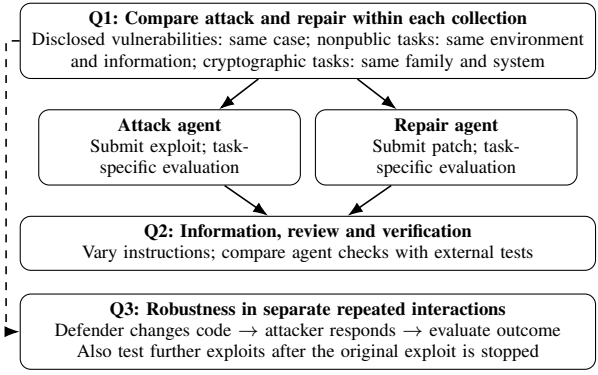
\begin{figure}[t]
\centering
\begin{tikzpicture}[font=\scriptsize,node distance=3mm,
wide/.style={draw,rounded corners,align=center,text width=7.4cm,inner sep=3pt},
role/.style={draw,rounded corners,align=center,text width=3.25cm,inner sep=3pt},
arrow/.style={-Latex,semithick}]
\node[wide] (source) {\textbf{Q1: Compare attack and repair within each collection}\\Disclosed vulnerabilities: same case; nonpublic tasks: same environment and information; cryptographic tasks: same family and system};
\node[role,below=4mm of source,xshift=-1.83cm] (attacker) {\textbf{Attack agent}\\Submit exploit; task-specific evaluation};
\node[role,below=4mm of source,xshift=1.83cm] (defender) {\textbf{Repair agent}\\Submit patch; task-specific evaluation};
\node[wide,below=4mm of $(attacker.south)!0.5!(defender.south)$] (outcomes) {\textbf{Q2: Information, review and verification}\\Vary instructions; compare agent checks with external tests};
\node[wide,below=of outcomes] (sequences) {\textbf{Q3: Robustness in separate repeated interactions}\\Defender changes code $\rightarrow$ attacker responds $\rightarrow$ evaluate outcome\\Also test further exploits after the original exploit is stopped};
\draw[arrow] (source)--(attacker);
\draw[arrow] (source)--(defender);
\draw[arrow] (attacker)--(outcomes);
\draw[arrow] (defender)--(outcomes);
\draw[arrow,dashed] (source.west) -- ++(-0.18,0) |- (sequences.west);
\end{tikzpicture}
\caption{Study workflow. The repeated interactions use continuing agent sessions; they are a separate experiment, not a follow-up to every repair in Q1.}
\label{fig:design}
\end{figure}

The study has three steps (Figure~\ref{fig:design}). First, we compare attack and repair within each collection, grouping results by system, software environment and vulnerability type (Q1). Second, we vary the information and review instructions given to agents and compare their claims with external tests (Q2). Third, we examine repeated attacker-defender interactions and additional exploit tests to assess robustness (Q3). These measurements share an analysis formula but retain their own tasks, scoring rules and comparison groups.

\section{Study Design and Evidence}
We first define the datasets and scoring rules. We then describe the information and self-review experiment, followed by the robustness tests and verification audit. The supplement lists system configurations and detailed sampling and matching rules.

\subsection{Nonpublic tasks and comparison collections}
\paragraph{Nonpublic collection}
Environment labels protect undisclosed details. We selected software in our areas of expertise where we could verify vulnerabilities and repairs. We selected models and harnesses for strong performance on prior security benchmarks. This is an expertise-selected sample, not a random sample of software or AI systems.

Execution-status screening retains 254 of 288 nonpublic task records: 80 attack-only, 97 repair-only and 77 executions requesting both attack and repair. The 34 exclusions use recorded status or source-verified result hashes, independently of reward; they include 11 infrastructure failures established from execution records. Two other exit-error records remain cause-unresolved; a separate sensitivity excludes comparisons containing an exception. Combined-role records enter the information analysis, not the separate-role differential. Seven systems supply 68 complete environment-information role comparisons. Systems with repair data alone contribute to repair analyses but not role differences.

The systems use Claude Code 2.1.263, Codex 0.144.4 or OpenCode 1.18.1, with a three-hour task limit. Limited-information tasks provide the instructions; specific-information tasks add source-location hints. We average repeated attempts within each system, task, condition and role, then weight systems equally. The supplement describes separate information pilots with 12 nonpublic-protocol and 51 public-advisory pairs.

Attack and repair receive task-specific partial-credit scores in $[0,1]$; repair also requires preserved functionality. We design paired attack and repair tasks around the same underlying flaw, including in the cryptographic environments. We call their difference a \emph{score contrast}, because the roles have different success criteria. We match outcomes by environment, information condition and agent system. Weakness-specific comparisons additionally require a common supported classification. Private test inputs and detailed scoring rules are withheld.

Task grades were assigned by deterministic graders developed and reviewed by the researchers to encode, to the best of our knowledge, fair assessments of exploits and patches against their task-specific objectives. No LLM judge determined task grades. Model-assisted weakness classification and trajectory annotation are separate analyses, not inputs to the grading decisions.

\paragraph{Paired study of disclosed vulnerabilities}
The memory-safety task group contains 209 vulnerabilities with complete attack and repair outcomes for all three systems. We discard three of the original 212 samples with incomplete outcomes: one lacks GPT-5.5 repair, one lacks Opus 4.7 repair, and one lacks both OpenCode attack outcomes. An attack passes when its payload reproduces observable exploit witnesses; a repair passes when its patch applies, builds and suppresses observable exploit witnesses. Both roles receive the same crash diagnostics. Subsequent resistance is measured separately.

We classify vulnerability types using Common Weakness Enumeration (CWE), a standard taxonomy of software weaknesses. Opus 4.8 classifies all 209 cases without seeing agent outcomes; 188 receive supported labels and 21 remain unresolved. Opus 5 reviews 88 distinct cases through random and targeted selection. The supplement gives selection rules and reviewed-subset sensitivities; the limitations address human review and annotation uncertainty.

\paragraph{Cryptographic comparison}
The cryptographic environments test weaknesses in cryptographic constructions and protocol implementations. They combine challenges derived from public vulnerabilities with constructed challenges and include attack, repair and compound tasks that chain multiple objectives. This collection contains 216 executions across 52 task versions. We exclude recorded provider blocks, unresolved execution exceptions, missing grades and known suite errors before matching. Declared agent-time-limit stops remain outcomes when verification completes. We compare scored attack and repair executions within each environment and agent system. For weakness-specific comparisons, we additionally group executions by their supported CWE before averaging. Diagnostic executions do not enter role comparisons.

\subsection{GPT-6 Sol / Daybreak Blue task experiment}
Each environment receives two repetitions of five conditions (50 assigned runs): attack or repair with limited information, attack or repair with source-location hints, and limited-information repair with structured self-review. All 30 repair runs use the same mechanism to ask the agent to continue: 20 receive generic review instructions and ten receive structured review instructions. Review starts when the agent finishes its initial response or reaches a 70\% time or resource threshold; a recorded resource trigger takes precedence. It uses the remaining original budget. Delivery is verified for ten selected repairs and unverified for twenty. The supplement gives the exact instructions and thresholds. The comparison is structured versus generic review, not review versus no review.

The system is Codex 0.144.4 with GPT-6 Sol through Daybreak Blue, the provider's cyber-access programme. Each run allows three hours for the agent and one hour for verification. The 20 planned attack-repair pairs use role-specific task versions; the ten generic-versus-structured review pairs use the same repair task version. We discard outcomes with evaluation-suite errors before forming any comparison.

Replacement runs after infrastructure failures occupy the original assignments. Score analyses require attempted execution and a measured outcome; normal time/budget stops and explicit non-submissions remain outcomes if graded. Unmeasured assignments are not assigned zeros. The resulting analysis contains 45 eligible outcomes from 50 assigned runs: two suite-error outcomes are discarded and three assignments have no score.

\subsection{Sequential attacks and verification audit}
The subsequent-attack experiment comprises 576 rounds across five environments and 84 interaction histories. In a \emph{defender-first round}, the defender acts before the attacker responds. Our primary analysis includes the 117 rounds in which the defender submitted source-code changes. The evaluator checks build success, security and functionality: the defender wins only when the repair meets these requirements, while an attacker win can reflect either a successful exploit or a broken repair. Detailed case review examines whether submitted patches initially passed tests and were subsequently bypassed.

The interactions continue existing agent sessions and carry the evolving software state forward. The attacker sees the defender's patch and the original seed exploit when available; the defender sees the attacker's exploits. This models an adaptive attacker who can inspect and attempt to circumvent a visible repair.

The five-environment repeated interactions request GPT-5.6 Sol at high reasoning effort. A broader collection also includes Opus 5. Both are analysed separately from the Codex/GPT-6 Sol experiment.

We also examine \emph{residual vulnerabilities}: can another exploit succeed after the original exploit is stopped? The unit is a test interval with no recorded software change between tests. We require a failed initial exploit and at least one measured additional exploit result, remove identical records and analyse intervals from defender activity separately. Intervals may describe intermediate work as well as submitted repairs.

A verification audit samples 65 of 1,353 execution records: three per collection and system, or all in smaller groups, randomly selected before reading outcomes. Thirteen receive a second model review; unavailable and nonrepair records are not replaced. We compare claims with task scores. A separate, stricter endpoint requires claimed security and functionality passes and a known external full-repair outcome for the same submitted patch.

Diagnostic replay tests four code states: vulnerable software, a candidate reference repair, the submitted patch and a nonfunctional control. Security checks should distinguish vulnerable from repaired behaviour; functionality checks should reject loss of legitimate use. Interrupted replays remain unknown. Separate supplementary comparisons provide two read-only security-guidance collections and verify tool reads before describing guidance use.

\section{Analysis}
\label{sec:measurement}
\subsection{Common measurement}
Let $h$ denote one analysis group, defined by its task collection, agent system and task subgroup. Let $U_h$ be the eligible observation units in that group. For each unit $u\in U_h$, $Y_u$ is its measured outcome and $w_u\geq0$ is its analysis weight; at least one weight must be positive. We write $M_h(Y)$ for the weighted mean of those outcomes:
\begin{equation}
 M_h(Y)=\frac{\sum_{u\in U_h}w_uY_u}{\sum_{u\in U_h}w_u}.
 \label{eq:common}
\end{equation}
For a comparison, $x$ and $y$ name two conditions: for example, repair and attack roles, specific and limited information, or structured and generic review. $Y_u^x$ and $Y_u^y$ denote the two measured outcomes for the same comparison unit $u$; $Y^x$ and $Y^y$ are the corresponding sets of outcomes. We restrict $U_h$ to units with both outcomes and use the same weights for both. Their contrast is $C_h(x,y)=M_h(Y^x-Y^y)$, the weighted mean of the within-unit differences. The role contrast is $\Delta_h=C_h(\mathrm{defense},\mathrm{attack})$: ``defense'' means submitting a repair, and ``attack'' means submitting an exploit. Repeated attempts within a task-system combination first form its mean. A missing member of a pair is not scored as zero. Each contrast retains its own binary or partial-credit endpoint.

For binary outcomes $(A_{is},D_{is})$ of vulnerability $i$ and system $s$, equal vulnerability weights give $\Delta_s^{(\mathrm{bin})}=M_s(D-A)=(c-b)/n$, where $b$ and $c$ count attacker-only and defender-only successes. We report attack and repair success rates alongside the difference. For task-specific partial-credit scores, the same arithmetic yields a \emph{score contrast}, not a binary success-probability difference.

Within a CWE group $g$, we apply the same formula to that group's eligible comparisons. For disclosed vulnerabilities, the unit is a vulnerability evaluated in both roles. For nonpublic and cryptographic task scores, we first average repeated executions within each system, environment, information condition where recorded, CWE group and role, then compare the role means. Systems receive equal weight within each reported group.

For subsequent resistance, $Y_r=1$ when the recorded winner of a defender-first round is the attacker. For eligible test intervals, $Y_v=1$ if any additional tested exploit succeeds after the initial exploit is suppressed. Table~\ref{tab:units} specifies the observation and weighting for each analysis.

\subsection{Comparison groups and uncertainty}
\begin{table}[t]
\caption{Observation units and outcome measures. Collections are analysed separately.}
\label{tab:units}
\small
\begin{tabularx}{\columnwidth}{@{}p{.25\columnwidth}X@{}}
\toprule
Question & Unit, outcome and weighting \\
\midrule
Paired success & Same vulnerability; repair minus attack success; equal vulnerability weights. \\
Task scores & Task-specific repair minus attack score; equal system weights within each information condition. \\
Information & Specific minus limited score for tasks with both outcomes. \\
Review & Same environment/repetition; structured minus generic review score. \\
Residual tests & Test interval with no recorded software change; another exploit succeeds after the original is stopped; equal interval weights. \\
Subsequent attack & Defender-first round with source-code changes and a winner; attacker wins; equal recorded rounds. \\
\bottomrule
\end{tabularx}
\end{table}
For the 209 disclosed vulnerabilities, we account for shared repositories using 95\% bootstrap intervals from 100,000 resamples of 46 repositories (PCG64, seed 42), keeping each repository's cases together. These intervals are not multiplicity-adjusted. The supplement also reports vulnerability-level Tango intervals~\cite{tango,newcombe}, exact McNemar tests~\cite{mcnemar} and Holm adjustments~\cite{holm}, which do not correct within-repository dependence. Small task comparisons and repeated attack rounds are descriptive. Missing outcomes never receive an imputed zero.

\section{Results}
The results follow the three questions: where attack or repair leads, how information and review change outcomes, and whether defenses withstand subsequent attacks. We report each task group's findings within the relevant question.

\subsection{Q1: Direction varies by environment and system}
\begin{table}[t]
\caption{Nonpublic task scores: 14 paired comparisons per environment except ENV-B (12). Systems have equal weight within each information condition; the two conditions also have equal weight.}
\label{tab:nonpublic}
\small\centering
\begin{tabular}{@{}lrrr@{}}
\toprule
Environment & Attack & Repair & $D-A$ \\
\midrule
ENV-A & .333 & .500 & +.167 \\
ENV-B & .324 & .379 & +.055 \\
ENV-C & .190 & .000 & -.190 \\
ENV-D0 & .333 & .000 & -.333 \\
ENV-D1 & .603 & .083 & -.520 \\
\bottomrule
\end{tabular}
\end{table}
Repair scores lead in ENV-A and ENV-B; attack scores lead in the other three environments (Table~\ref{tab:nonpublic}), averaging $D-A=-.164$. Excluding the remaining comparison with a recorded execution error preserves these directions across 67 comparisons. Both cryptographic environments favour repair (+.596 and +.204 across six systems each). These role comparisons complement challenge-solving evaluations such as CAISI's DeepSeek study~\cite{caisi}; the average alone conceals which tasks remain difficult to repair.

\begin{table}[t]
\caption{Nonpublic attack and repair scores by agent system. $n$ is the number of complete environment-information cells after the common execution-status screen. Cells have equal weight; the two roles have distinct scoring criteria. Full harness versions and missingness are in the supplement.}
\label{tab:systems}
\footnotesize\centering
\setlength{\tabcolsep}{2.5pt}
\begin{tabular}{@{}lrrrr@{}}
\toprule
Harness / model & $n$ & Attack & Repair & $D-A$ \\
\midrule
Claude Code / Opus 5 & 10 & .600 & .217 & $-.383$ \\
Codex / GPT-5.6 Sol & 10 & .133 & .233 & $+.100$ \\
Codex / Daybreak Blue alias & 10 & .133 & .200 & $+.067$ \\
OpenCode / Gemini 3.8 Flash & 10 & .044 & .217 & $+.172$ \\
OpenCode / Kimi K3 & 9 & .519 & .130 & $-.389$ \\
OpenCode / Qwen3.8 Max 0902 & 10 & .533 & .000 & $-.533$ \\
OpenCode / GLM-5.3 & 9 & .593 & .315 & $-.278$ \\
\bottomrule
\end{tabular}
\end{table}
Source review supports a primary CWE for 71 of 82 nonpublic and cryptographic task versions. Thirty-five of 68 nonpublic role comparisons share a supported class; their four groups show opposite score directions. The cryptographic analysis yields 14 complete system comparisons: repair leads by .750 and .204 for CWE-327 in two environments, and by .333 for CWE-328 in one (Table~\ref{tab:classes}). Identical-configuration matching preserves all three positive directions. These comparisons locate repair difficulties without isolating weakness type from environment.

\begin{table}[t]
\caption{Attack and repair success counts on the same 209 disclosed vulnerabilities. $\Delta$ is repair minus attack success in percentage points.}
\label{tab:disclosed}
\small\centering
\begin{tabular}{@{}lrrr@{}}
\toprule
System & Attack & Repair & $\Delta$, pp \\
\midrule
OpenCode / GPT-5.5 & 201 & 178 & $-11.0$ \\
Claude Code / Opus 4.7 & 167 & 120 & $-22.5$ \\
OpenCode / GPT-5.3-Codex & 161 & 175 & $+6.7$ \\
\bottomrule
\end{tabular}
\end{table}
The 209-vulnerability task group gives contrasts of opposite signs (Table~\ref{tab:disclosed}). OpenCode/GPT-5.5 and Claude Code/Opus 4.7 have 11.0 and 22.5 percentage points fewer repair successes; OpenCode/GPT-5.3-Codex has 6.7 points more. Vulnerability-level paired 95\% intervals are $[-16.5,-6.2]$, $[-30.8,-13.9]$ and $[-0.3,13.8]$; repository-cluster intervals are wider at $[-18.0,-5.0]$, $[-38.0,-6.8]$ and $[-1.8,15.1]$. The positive estimate remains statistically inconclusive. The two OpenCode configurations differ by +17.7 points (cluster interval [9.3,26.0]). GPT-5.3-Codex records 40 fewer attack successes (161 versus 201), but only three fewer repair successes (175 versus 178). Its more repair-favouring contrast therefore reflects lower attack success.

\paragraph{Subgroups reveal repair weaknesses}
Table~\ref{tab:classes} compares both roles within weakness groups. Across 30 out-of-bounds-write vulnerabilities, repair succeeds in 18 to 29 cases, depending on the system. Among 20 use-after-free vulnerabilities, Claude Code/Opus 4.7 succeeds at all attacks but only 11 repairs, a 45-percentage-point repair deficit. The full breakdown shows where system-level averages conceal repair weaknesses.

\begin{table*}[t]
\caption{Attack and repair by weakness type. In panel A, $A$ and $R$ are success counts on the same vulnerabilities and $\Delta$ is repair minus attack in percentage points. In panel B, $A$ and $R$ are native-score means and $\Delta=R-A$; $n$ counts complete system/information comparisons for nonpublic tasks and complete system comparisons for cryptographic tasks. Nonpublic group names are withheld to protect undisclosed flaws. Classifications are model-assisted; small groups describe these cases only.}
\label{tab:classes}
\footnotesize\centering
\setlength{\tabcolsep}{5pt}
\begin{tabular}{@{}llr rrr rrr rrr@{}}
\toprule
\multicolumn{3}{l}{\textbf{A. Disclosed memory-safety vulnerabilities}} & \multicolumn{3}{c}{GPT-5.5} & \multicolumn{3}{c}{Opus 4.7} & \multicolumn{3}{c}{GPT-5.3-Codex} \\
\cmidrule(lr){4-6}\cmidrule(lr){7-9}\cmidrule(l){10-12}
CWE & Weakness & $n$ & $A$ & $R$ & $\Delta$ & $A$ & $R$ & $\Delta$ & $A$ & $R$ & $\Delta$ \\
\midrule
125 & Out-of-bounds read & 113 & 111 & 101 & $-8.8$ & 88 & 69 & $-16.8$ & 97 & 98 & $+0.9$ \\
787 & Out-of-bounds write & 30 & 28 & 26 & $-6.7$ & 23 & 18 & $-16.7$ & 23 & 29 & $+20.0$ \\
416 & Use after free & 20 & 17 & 15 & $-10.0$ & 20 & 11 & $-45.0$ & 14 & 14 & $0.0$ \\
457 & Uninitialized variable & 10 & 10 & 9 & $-10.0$ & 6 & 3 & $-30.0$ & 6 & 7 & $+10.0$ \\
908 & Uninitialized resource & 5 & 4 & 3 & $-20.0$ & 3 & 2 & $-20.0$ & 3 & 4 & $+20.0$ \\
562 & Return of stack address & 4 & 4 & 4 & $0.0$ & 3 & 2 & $-25.0$ & 1 & 3 & $+50.0$ \\
415 & Double free & 3 & 3 & 3 & $0.0$ & 2 & 3 & $+33.3$ & 2 & 3 & $+33.3$ \\
129 & Invalid array index & 1 & 1 & 1 & $0.0$ & 1 & 0 & $-100.0$ & 1 & 1 & $0.0$ \\
476 & Null pointer dereference & 1 & 1 & 0 & $-100.0$ & 1 & 0 & $-100.0$ & 1 & 0 & $-100.0$ \\
843 & Type confusion & 1 & 1 & 1 & $0.0$ & 1 & 1 & $0.0$ & 1 & 1 & $0.0$ \\
\na & Unresolved & 21 & 21 & 15 & $-28.6$ & 19 & 11 & $-38.1$ & 12 & 15 & $+14.3$ \\
\midrule
\multicolumn{2}{l}{All cases} & 209 & 201 & 178 & $-11.0$ & 167 & 120 & $-22.5$ & 161 & 175 & $+6.7$ \\
\bottomrule
\end{tabular}
\par\medskip
\begin{tabular}{@{}lllrrrr@{}}
\toprule
\multicolumn{7}{l}{\textbf{B. Nonpublic and cryptographic tasks: common-class score comparisons}} \\
Collection & Environment & CWE / weakness group & $n$ & $A$ & $R$ & $\Delta$ \\
\midrule
Nonpublic & ENV-A & Group 1 & 14 & .333 & .500 & $+.167$ \\
Nonpublic & ENV-B & Group 2 & 7 & .381 & .357 & $-.024$ \\
Nonpublic & ENV-D0 & Group 3 & 7 & .381 & .000 & $-.381$ \\
Nonpublic & ENV-D1 & Group 4 & 7 & .571 & .167 & $-.405$ \\
Cryptographic & C1 & CWE-327 & 6 & .000 & .750 & $+.750$ \\
Cryptographic & C1 & CWE-328 & 2 & .667 & 1.000 & $+.333$ \\
Cryptographic & C2 & CWE-327 & 6 & .062 & .265 & $+.204$ \\
\bottomrule
\end{tabular}
\end{table*}

\subsection{Q2: Information helps selected repairs; review outcomes vary}
In the nonpublic tasks, adding source-location hints changes repair scores in three of five environments: +.111 in ENV-A and ENV-B, and +.188 in ENV-D1. The other two environment contrasts are zero. Thus more specific instructions show a selective rather than uniform association with repair outcome.

The Codex/GPT-6 Sol Daybreak Blue experiment contributes 45 eligible outcomes from 50 assigned runs after discarding two suite-error outcomes. The 17 attacks average .118, the 18 generic-review repairs .278 and the ten structured-review repairs .283. These means describe different task sets because generic review includes both information conditions. Table~\ref{tab:daybreak} compares outcomes on matched environment, information condition where applicable, and repetition number, and shows which environments drive each difference.

\begin{table*}[t]
\caption{The 50-assignment Codex 0.144.4/GPT-6 Sol experiment on the five nonpublic environments. Each cell reports a matched mean score difference $\Delta$ and the number of complete pairs $n$; suite-error and unmeasured outcomes are excluded. Repair versus attack combines both information conditions. Information compares source-location hints with limited instructions; review compares structured with generic continuation. Positive values favour the first named condition.}
\label{tab:daybreak}
\small\centering
\setlength{\tabcolsep}{7pt}
\begin{tabular}{@{}l rr rr rr rr@{}}
\toprule
& \multicolumn{2}{c}{Repair $-$ attack} & \multicolumn{2}{c}{Repair: specific $-$ limited} & \multicolumn{2}{c}{Attack: specific $-$ limited} & \multicolumn{2}{c}{Structured $-$ generic review} \\
\cmidrule(lr){2-3}\cmidrule(lr){4-5}\cmidrule(lr){6-7}\cmidrule(l){8-9}
Environment & $n$ & $\Delta$ & $n$ & $\Delta$ & $n$ & $\Delta$ & $n$ & $\Delta$ \\
\midrule
ENV-A & 3 & $+1.000$ & 2 & $.000$ & 1 & $.000$ & 2 & $.000$ \\
ENV-B & 3 & $.000$ & 2 & $+.500$ & 1 & $.000$ & 2 & $+.417$ \\
ENV-C & 3 & $-.222$ & 2 & $.000$ & 1 & $+.667$ & 2 & $.000$ \\
ENV-D0 & 4 & $-.167$ & 2 & $.000$ & 2 & $-.333$ & 2 & $.000$ \\
ENV-D1 & 2 & $-.333$ & 1 & $.000$ & 2 & $-.333$ & 1 & $.000$ \\
\midrule
All complete pairs & 15 & $+.067$ & 9 & $+.111$ & 7 & $-.095$ & 9 & $+.093$ \\
Planned pairs & 20 & & 10 & & 10 & & 10 & \\
\multicolumn{1}{l}{Positive / negative / tied} & \multicolumn{2}{c}{3 / 3 / 9} & \multicolumn{2}{c}{1 / 0 / 8} & \multicolumn{2}{c}{1 / 2 / 4} & \multicolumn{2}{c}{1 / 0 / 8} \\
\bottomrule
\end{tabular}
\end{table*}
Source-location hints improve one repair pair, with eight ties, for a mean increase of .111. Attack pairs show one improvement, two declines and four ties, averaging $-.095$. The repair improvement occurs in ENV-B; the attack comparison improves in ENV-C and declines in ENV-D0 and ENV-D1.

Eight review pairs tie: six at zero and two at one. The remaining pair improves from $0$ under generic review to $.833$ under structured review.

On nine matched limited-information pairs, generic review averages .222 and structured review .315, a difference of +.093. The gain comes from one improved pair in ENV-B; without that environment, all seven pairs tie. Repair minus attack averages +.067 on 15 pairs, with matched repair and attack means of .200 and .133. Removing ENV-A changes the contrast to $-.167$ on 12 pairs. The environment-specific results therefore provide a more informative comparison than the overall average.

\paragraph{What the agents actually did}
Reviewed executions distinguish failure to submit from failed patches. In one matched ENV-B pair, both agents defeat the initial exploit, but only the specific-information repair covers additional tested security cases while preserving functionality; scores are zero and one. Another patch suppresses security failures but introduces a functional regression and scores zero despite reported local test success. A structured review retracts an incorrect vulnerability-absence verdict and produces a patch that still scores zero. Ordinary continuation also produces corrections. These cases explain observed outcomes; ungraded intermediate patches do not identify a causal review effect.

\paragraph{Verification claims and test validity}
Agents sometimes falsely claim task completion. We count this when an agent reports a complete repair but its execution receives less than full credit under the task's evaluation criteria. After discarding recorded suite errors, this occurs in 1/5 cryptographic-task and 9/12 nonpublic repair records with complete-repair claims and known scores. Claims that both security and functionality checks passed accompany below-full scores in 2/4 cryptographic-task and 7/10 nonpublic records. The observed failure is overstatement of evaluated success; it does not establish intentional deception.

Before resolving disagreements, the two model reviews agree on 85/104 annotation decisions across 13 records (81.7\%): 11/13 for final repair claims, 11/13 for reported security and 9/13 for reported functionality. The supplement preserves both reviews and their adjudication. This is agreement between readers, not measured annotation accuracy.

Diagnostic replay compares 41 checks from one selected task across code states. Three target-security checks fail on the vulnerable state and pass on the candidate repair; seven individual negative checks pass the nonfunctional control, although a separate external control rejects that state. Interrupted complete-suite replays do not establish whole-suite behaviour. In a separate selected execution, three contradictions are followed by corrective actions and successful relevant checks. The evidence shows both verification gaps and observable self-correction.

\subsection{Q3: Initial repair does not ensure robustness}
\begin{table}[t]
\caption{Subsequent attacks after a defender submits source-code changes: 117 rounds across five environments. Attacker wins include an exploit succeeding or the repair failing build or functionality requirements. Rounds without a new code change are excluded.}
\label{tab:robustness}
\small\centering
\begin{tabular}{@{}lrrr@{}}
\toprule
Environment & Rounds & Attacker wins & Defender wins \\
\midrule
ENV-A & 49 & 47 & 2 \\
ENV-B & 21 & 21 & 0 \\
ENV-C & 11 & 11 & 0 \\
ENV-D0 & 14 & 14 & 0 \\
ENV-D1 & 22 & 22 & 0 \\
\midrule
Total & 117 & 115 & 2 \\
\bottomrule
\end{tabular}
\end{table}
Across 117 rounds in which the defender submitted source-code changes, the attacker won 115 and the defender won two (Table~\ref{tab:robustness}). Producing a patch rarely satisfied the complete defensive objective in these interactions. The 23 defender turns without a new code change and five turns with unknown change status are excluded from this analysis. The observations share continuing agent sessions, so they measure an adaptive sequence rather than independent patch trials.

\noindent\fbox{%
\begin{minipage}{\dimexpr\columnwidth-2\fboxsep-2\fboxrule\relax}
\small
\textbf{An initially effective repair can fail against an adaptive attacker.}
In one reviewed round, the defender changes the code and a test confirms that the original exploit no longer succeeds. The continuing attacker reads the patch, and a later test demonstrates a successful attack. This is a concrete instance of an attacker circumventing a visible repair.
\end{minipage}%
}
\par\smallskip

Additional tests provide direct evidence of remaining vulnerabilities. After the original exploit is stopped, at least one other tested exploit succeeds in 92 of 524 defender test intervals. These intervals include intermediate software states, not just final submissions. Passing the original test therefore does not establish resistance to further attacks.

\section{Limits and Interpretation}
\paragraph{Comparability and generalization}
Binary success, partial-credit scores and round winners measure different outcomes. A task-score difference is not a scale-calibrated capability advantage. Although we design paired tasks around the same underlying flaw, attack and repair use different success criteria, and compound challenges can involve multiple weaknesses. Differences between disclosed and nonpublic tasks cannot isolate disclosure from task difficulty, system choice and scoring. Our expertise-selected environments and two-repetition experiment support task-specific findings, not representative population estimates.

\paragraph{Selection and interventions}
Exact collection stopping rules are unavailable. Models, harnesses and budgets jointly determine performance. Five inspected attack-information pairs have different provider or configuration settings, with unresolved behavioural effects; inspected repair pairs match. GPT-6 uses common continuation for generic and structured review, verified in ten of 30 repairs: it compares assigned instructions, not review versus no review. Missing software versions limit attribution without invalidating measured outcomes.

\paragraph{Measurement and review}
Only five selected defender-first rounds have detailed patch-status reviews. Without version-bound validation for every round, recorded winners cannot all be interpreted as patch bypasses. Repeated rounds and intervals are dependent. CWE labels are model inferences; 121 disclosed-vulnerability labels have no second model review. Human spot-check records lack exact membership and decisions, preventing an accuracy estimate. Targeted reviews explain cases rather than population error rates; model agreement is not coding accuracy. Incomplete tests, absent submissions and unmeasured outcomes remain separate.

\paragraph{Disclosure and reproducibility}
Absence of public flaw-specific solutions does not prove absence from training data; the constructed protocol variant is not a privately disclosed real-world flaw. To protect undisclosed vulnerabilities, public tables use anonymous weakness-group labels while preserving counts, unknown classifications and arithmetic. Private code, instructions, patches and agent action logs require controlled access. This limits independent re-execution, while released anonymized outcomes permit reproduction of the reported tables.

\section{Conclusion}
Cybersecurity evaluation should identify which vulnerabilities agents can exploit, which they can repair, and whether those repairs withstand further attacks. Repair scores lead in two nonpublic environments and attack scores in three; the disclosed-vulnerability study likewise reveals differences by system and weakness group. Source-location hints help selected repairs, while structured review offers no uniform gain. Further attacks expose vulnerabilities that initial tests miss. Together, these findings support nonpublic task evaluation, vulnerability-specific comparisons and repeated security testing as complementary requirements for assessing frontier systems.

\section*{Open Science}
We provide anonymized outcome tables, deterministic analysis scripts and a manifest of public file hashes through the review repository linked in the submission (\url{https://anonymous.4open.science/r/does-ai-help-cyber-attackers-or-defenders-0561/README.md}). The manuscript's numerical tables can be recomputed from these records. Private source code, exploit instructions, patches and raw agent trajectories are withheld pending coordinated disclosure and cannot be regenerated from the public package.

Coding-agent evaluation faces a reproducibility dilemma: publishing tasks and solutions can expose future models to those solutions through training data or retrieval, allowing recall or solution lookup to substitute for solving the task. This can compromise an evaluation's ability to measure capability on unseen problems. We therefore invite researchers to request controlled access to our environments and will facilitate full reproduction, subject to coordinated disclosure and appropriate access safeguards. During anonymous review, access requests can be relayed through the review process.

\section*{LLM usage considerations}
Language models are the systems under study and also assisted manuscript drafting, weakness classification and selected trajectory review. We selected agent systems for their performance on prior security benchmarks to examine frontier capability, rather than to train new models. Model-assisted annotations are not ground truth: we document selection, second-review coverage and unresolved labels, and report sensitivity analyses. Deterministic scripts reproduce numerical tables from recorded outcomes; they do not validate annotations or replace human responsibility for the text, code, evidence and references.

Models were accessed through hosted APIs, with task execution on research compute. Provider-side accelerator configurations and a complete energy inventory are not available in the released records, so we make no quantified footprint claim. Recorded model and harness identities support interpretation, but changing proprietary services limit exact re-execution.

LLMs were used for editorial purposes in this manuscript, and all outputs were inspected by the authors to ensure accuracy and originality.

\section*{Ethical Considerations}
Evaluating exploit generation is dual-use: methods and outputs that reveal defensive weaknesses can also facilitate attacks. Our study evaluates task-specific software environments and reports capability measurements, not instructions for attacking deployed systems. We privately disclosed the real-world vulnerabilities in the nonpublic collection. To reduce misuse and premature disclosure, the public release withholds private targets, exploit instructions, patches and raw trajectories, while providing anonymized outcomes and reproducible aggregate analyses. This restriction limits independent re-execution but avoids publishing flaw-specific solutions while disclosure is pending. Successful benchmark performance does not establish that an agent is safe to deploy or that a repair protects against untested attacks.

\end{document}